\documentclass[11pt,onecolumn,floatfix,superscriptaddress,nofootinbib]{revtex4-1}

\usepackage{graphicx}                   
\usepackage{epsfig}
\usepackage{amsmath,amssymb}            
\usepackage{hyperref}                   
\usepackage{epstopdf}
\usepackage{placeins}
\usepackage{upgreek}
\usepackage{slashed}
\usepackage{color}
\usepackage{footmisc}
\usepackage{footnote}
\usepackage{mathrsfs}
\definecolor{orange}{rgb}{1,0.5,0}
\definecolor{brown}{rgb}{0.65, 0.16, 0.16}
\definecolor{phlox}{rgb}{0.87, 0.0, 1.0}

\graphicspath{{figs/}}                  

\begin{document}

   \title{Muon anomalous magnetic moment and Right handed sterile neutrino }

\author{Iman Motie}
\email{iman.motie@univ-tlse3.fr}
\affiliation{Universitè de Toulouse, UPS-OMP, IRAP, F-31400 Toulouse, France,\\
     CNRS, IRAP, 14 avenue Edouard Belin, F-31400 Toulouse, France }	
\author{Mahdi Sadegh}
\email{m.sadegh@hormozgan.ac.ir}
\affiliation{Minab Higher Education Center, Hormozgan University, P. O. Box: 3995, Bandar Abbas, Iran}
\author{S. Mahmoudi}
\email{s.mahmoudi@ipm.ir}
\affiliation{School of Particles and Accelerators, Institute for Research in Fundamental
	Sciences (IPM), P. O. Box 19395-5531, Tehran, Iran}
\author{Jafar Khodagholizadeh}
\email{gholizadeh@ipm.ir}
\affiliation{Farhangian University, P.O. Box 11876-13311, Tehran, Iran}
\author{Alain Blanchard}
\email{alain.blanchard@irap.omp.eu }
\affiliation{Universitè de Toulouse, UPS-OMP, IRAP, F-31400 Toulouse, France,\\
     CNRS, IRAP, 14 avenue Edouard Belin, F-31400 Toulouse, France }
\author{She.-Sheng. Xue}
\email{xue@icra.it}
\affiliation{
ICRANet Piazzale della Repubblica, 10 -65122, Pescara, Italy,
\\
Physics Department, Sapienza University of Rome, 
Rome, Italy,\\
 INFN, Sezione di Perugia, 
Perugia, Italy,
\\
 ICTP-AP, University of Chinese Academy of Sciences, Beijing, China.
}
      
    \begin{abstract}
    The muon's magnetic moment is a fundamental quantity in particle physics and the observed deviation of its value from the prediction of quantum electrodynamics (QED) motivated investigations beyond the Standard Model (SM). In this study, we utilize the effective coupling of right-handed sterile neutrinos with SM gauge bosons to calculate the muon anomalous magnetic moment ($\boldsymbol{\mu}$AMM) at one-loop level. The contribution of the sterile neutrino interactions on the $\boldsymbol{\mu}$AMM is evaluated by including both standard and non-standard neutrino interactions. Our analysis shows that the standard sterile neutrino interactions give a negligible contribution to $\Delta a_{\boldsymbol{\mu}}$ while the non-standard neutrino interactions can play a significant role in explaining the $(g-2)_{\boldsymbol{\mu}}$ anomaly. In the context of the non-standard neutrino interaction, our calculation shows that a Dirac mass scale $M_D\approx100\,\text{GeV}$ could explain the muon anomaly if the right handed sterile neutrino's coupling with SM particles is about $\mathcal{G}_R\approx 10^{-3}$. Furthermore, we use the observed discrepancy in the muon anomalous magnetic moment to impose constraints on the model parameters. We present the allowed parameter space, consistent with the experimental data on $\Delta a_{{\boldsymbol{\mu}}}^{SN}$ and discuss the percentage of the ${\boldsymbol{\mu}}$ anomaly compensation in terms of the coupling constant $\mathcal{G}_R$.  
      
    \end{abstract}


\maketitle

   \section{Introduction}
 Over the past decades, despite the great success of the SM of particle physics, it
still challenged by some experimental anomalies, such as dark matter (DM) relic density \cite{Green:2021jrr, Roszkowski:2004jc}, mass of neutrinos \cite{SNO:2002hgz}, and baryon asymmetry of the Universe \cite{Canetti:2012zc}.  These enduring anomalies suggest that the SM is incomplete and new physics beyond the SM may be required to address these issues. \par


Another long-standing anomaly is related to the $\boldsymbol{\mu}$AMM, $a_{\boldsymbol{\mu}}\equiv (g-2)_{\boldsymbol{\mu}}/2$. More recently, it has been reported a new measurement of $a_\mu$ using data collected in 2019 (Run-2) and 2020 (Run-3) by the Muon $(g-2)$ Experiment at Fermi National Accelerator Laboratory (FNAL) \cite{Muong-2:2023cdq} as well as the Brookhaven National Laboratory (BNL) \cite{Muong-2:2006rrc}
\begin{equation}
\begin{cases}
a_{\boldsymbol{\mu}}^{{\scriptsize \mathrm{FNAL}}} = \left( 116 \ 592 \ 055 \pm 24 \right) \times 10^{-11}, \\\\
a_{\boldsymbol{\mu}}^{{\scriptsize \mathrm{BNL}}} = \left( 116 \ 592 \ 089 \pm 63 \right) \times 10^{-11}, 
\end{cases}
\end{equation} 
leading to the combined (BNL and FNAL) experimental average \cite{Muong-2:2023cdq}
\begin{align}
a_{\boldsymbol{\mu}}^{exp} = \left( 116 \ 592 \ 059 \pm 22 \right) \times 10^{-11},
\end{align}
which differs notably from the SM theory prediction \cite{Aoyama:2020ynm}
 \begin{align}
 	a_{\boldsymbol{\mu}}^{{\scriptsize \mathrm{SM}}} = \left( 116 \ 591 \ 810 \pm 43 \right) \times 10^{-11}.
 \end{align}
The above results represent that the SM prediction is more than $5\sigma$ smaller than the latest experimental measurement
\begin{align}
\label{exp-sm}
\Delta a_{\boldsymbol{\mu}}=a_{\boldsymbol{\mu}}^{\text{exp}}-a_{\boldsymbol{\mu}}^{\text{SM}}= (2.49 \pm 0.48) \times10^{-9}.
\end{align}
Although such a discrepancy makes the importance of clarifying the correct SM theoretical calculation significant \cite{Borsanyi:2020mff}, it has generated great interest in the particle physics community since \textcolor{black}{ if the current anomaly is confirmed,} it can \textcolor{black}{probably} be served as a strong evidence of new physics beyond the SM and might arise from the effects
of as-yet-undiscovered particles contributing through virtual loops.\par

Theoretical efforts to account for this discrepancy have included a variety of scenarios; More recent lattice QCD results attempt to explain $\mu$AMM by concentrating on enhancing the SM calculations with incorporating hadronic contributions more  accurately \cite{Borsanyi:2020mff}. However, there have been a great number of theoretical works trying to interpret the observed anomaly within the framework of beyond the SM \cite{Borah:2021jzu, Zu:2021odn, Athron:2021iuf, Keshavarzi:2021eqa, Lindner:2016bgg, Cao:2021tuh, Hammad:2021mpl, Aghababaei:2017bei, Dey:2021pyn, Chakraborti:2020vjp, Alvarado:2021nxy, Ghorbani:2021yiw, Anchordoqui:2021llp, Zhou:2021vnf, Lu:2021vcp}.
\color{black}
One of the most important avenues of these researches focus on the new physics models including sterile neutrinos. For instance, the $\mu$AMM has been discussed within the context of a theoretical framework based on the $SU_{L}(2)\times
SU_{R}(2) \times U(1)_{B-L}$ -gauge group in \cite{Boyarkin:2008zz}. In addition, this anomaly has been addressed through models with mirror symmetry and type I see-saw mechanism at low energy scale of electroweak interactions in \cite{Dinh:2023vqb}. Furthermore, Abdallah et al. in \cite{Abdallah:2011ew} have analyzed the $\mu$AMM in TeV scale $B-L$ extension
of the SM with inverse seesaw mechanism. \par

It is important to note that some of the most influential studies in recent years are those that explore a possible link between the muon anomalous magnetic moment anomaly and the origin of neutrino masses. For instance, authors in \cite{Majumdar:2020xws, Zhou:2021vnf} have considered a gauged $U(1)_{L_{\mu}-L_{\tau}}$ extension of the left-right symmetric theory in order to
simultaneously address neutrino mass, mixing and the $\mu$AMM via a light $Z'$ boson. Additional prominent examples include radiative neutrino mass models which introduce new charged scalars contributing to both neutrino masses and 
$\mu$AMM \cite{Chakrabarty:2018qtt, Barman:2021xeq}. Furthermore, models 
 featuring leptoquarks~\cite{Nomura:2021oeu, Zhang:2021dgl, Chen:2022hle}, also attempt to unify the explanation of neutrino mass generation and the $\mu$AMM discrepancy within a common framework.

\par
\color{black}
In this work, we continue the investigation of the muon anomaly in the context of an effective model
based on the fundamental symmetries and particle content of the SM \cite{Xue:2016dpl, Xue:2016txt}. 
This scenario was motivated by the parity symmetry reconstruction at high energies without any extra gauge bosons and
incorporates three massive sterile neutrinos $\nu_{R}$ as well as the SM gauge symmetric four-fermion interactions giving
rise to new effective interactions between sterile neutrinos and SM gauge bosons. We will study
the interaction effects of this sort of sterile neutrino with SM particles on the ${\boldsymbol{\mu}}$AMM. In addition, we will employ the observed discrepancy in $(g-2)_{\mu}$ to set constraint on the model parameters, such as coupling constant.\par

This paper is organized as follows. A brief review of the theoretical aspects of the $g-2$ has been provided in Section \ref{review}. In Section \ref{sec3}, we review the sterile neutrino`s effective lagrangian and its interaction with gauge bosons. Section \ref{sec4} is devoted to the calculation of the sterile neutrino corrections to ${\boldsymbol{\mu}}$AMM: In subsection \ref{sec4-1}, we study the contribution due to $W$ boson and $\nu_{R}$ mediation (standard interaction) and the contribution arising from $W$ boson mediation with Left-Right neutrino mixing (non-standard interaction) is discussed in subsection \ref{sec4-2}. In Section \ref{secsumm}, we give some concluding remarks. Finally, the detail of calculations of the sterile neutrino corrections to $a_{\boldsymbol{\mu}}$ has been presented in Appendix \ref{appe} and Appendix \ref{appe2}.

\color{black}
  \section{Electromagnetic form factors and $\mu$AMM }\label{review}
According to quantum mechanics, any elementary charged particle with intrinsic angular momentum ($\vec{s}$) has a magnetic dipole
moment ($\vec{\mu}$), which is related to its spin through the following equation, 
\begin{equation}
	\vec{\mu} = g \left( \frac{q}{2m}\right) \vec{s},
\end{equation}
where $q =\pm e$ is the electric charge of a given charged particle and $m$ denotes its mass. The factor $g$ is the gyromagnetic ratio which equals to 2 in Dirac theory. However, calculations of the loop corrections in quantum field theories, like the SM, indicate that this quantity receives contributions from radiative corrections. Indeed, since  the interaction of the elementary particle with a photon
is modified by additional interactions with virtual particles, the value of $g$ is modified, resulting in increasing its value from the tree-level prediction of $g = 2$. For the charged lepton ($l = e, {\boldsymbol{\mu}}, \tau$), these corrections are parametrized in terms of $a_l$ which is defined as the fractional deviation from the tree-level prediction of $g_{l} = 2$:
\begin{equation}
	a_{l}=(g-2)_{l}/2,
\end{equation}
referred to as the anomalous magnetic moment. This quantity has played the key role for years, serving as one of the most accurate ways to test of the SM. To date, a great deal of effort has been devoted to determine the SM modifications to $g$ from virtual SM particles up to a sufficient order  ~\cite{Aoyama:2020ynm, Borsanyi:2020mff, Burnett:1967zfb, Stoffer:2023gba, Lautrup:1968tdb, Aldins:1970id, Lautrup:1972iw, PhysRevD.5.2269}. The comparison between the theoretical predictions and experimental measurements of $a_l$ leads to the studies of lepton magnetic moments which is a powerful indirect probe for new physics. 
\begin{figure}[t]
	\centering
	\includegraphics[scale=0.9]{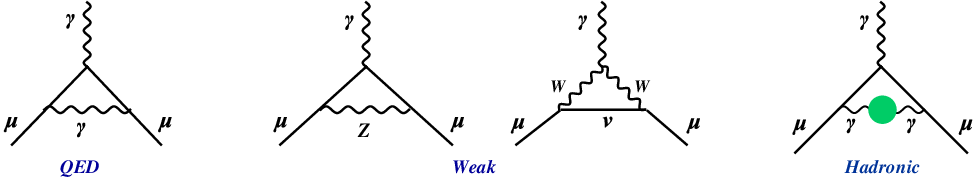}
	\caption{Lowest-order SM corrections to $a_{\mu}$. From left to right: QED, weak, and hadronic \cite{ Lindner:2016bgg}.} 
	\label{fig:1}
\end{figure} 

Regarding the muon particle, there exists a deviation between the prediction of the SM and the most precise measurement performed to date\cite{Aoyama:2020ynm, Muong-2:2023cdq}. Additionally, the experimental value of $a_{\boldsymbol{\mu}}$ cannot be explained solely by the SM, and one must take into account not only the loop corrections through the SM framework but also the contribution from new physics beyond the SM \cite{smgfactor1,smgfactor2,Aghababaei:2017bei}.
\begin{eqnarray}
	a_{\boldsymbol{\mu}}=a_{\boldsymbol{\mu}}^{SM} + a_{\boldsymbol{\mu}}^{NP},
\end{eqnarray}
where $a_{\boldsymbol{\mu}}^{NP}$ contains all effects on the anomalous magnetic moment of the muon from new physics beyond the SM and  
\begin{eqnarray}
a_{\boldsymbol{\mu}}^{SM}=a_{\boldsymbol{\mu}}^{QED}+a_{\boldsymbol{\mu}}^{Weak}+a_{\boldsymbol{\mu}}^{hadron},
\end{eqnarray}
where the $a^{QED}_{\boldsymbol{\mu}}$  includes the Schwinger result \cite{PhysRev.73.416,PhysRev.74.250} plus corrections up to five loops, $a^{Weak}_{\boldsymbol{\mu}}$ shows the weak contribution with the loops containing the heavy bosons $W^{\pm}$, Z, and H and the hadronic part  $a^{hadron}_{\boldsymbol{\mu}}$ shows the contribution of hadrons in the loop corrections\cite{Aoyama:2020ynm}. The relevant Feynman diagrams are depicted in Fig.~\ref{fig:1}. \par

In quantum electrodynamics (QED), loop effects in the electromagnetic interaction of fermionic point particles are typically studied using electromagnetic form factors to parameterize currents. The vector current is a Lorentz vector that can be expanded in terms of all independent Lorentz vectors in the system being considered. This means that the most general form for the electromagnetic current between Dirac leptons, satisfying Lorentz covariance and the Ward identity, can be expressed as follows
\begin{eqnarray}
	<\psi(p)|J^{EM}_\mu|\psi(p')>&=& \bar{u}(p'){\cal{F_\mu}} (q^2)u(p),
\end{eqnarray}
where $q_\mu = p'_\mu-p_\mu$ and
\begin{eqnarray}
	{\cal{F_\mu}}(q^2)&=&F_1\,\gamma_\mu+F_2\,\,i\frac{\sigma_{\mu\nu}q^\nu}{2m}
	+F_3(q_\mu-\frac{q^2}{2m}\gamma_\mu)\gamma_5
	+F_4\,\, \sigma_{\mu\nu}
	\frac{q^\nu}{2m}\gamma_5,\label{form-factor}
\end{eqnarray}
in which, $m$ stands for the mass of the charged lepton (specifically the muon), while $F_i$'s with $i$ ranging from 1 to 4 represent the standard electric charge, magnetic dipole, anapole (axial charge), and electric dipole form factors, respectively. 
In this study, we focus exclusively on $F_2$ for muon and compute its corrections from the right handed sterile neutrinos' contribution in the following sections.

\section{Sterile Neutrinos interactions with SM particles}
\label{sec3}

Sterile neutrinos are a type of hypothetical particle that could potentially elucidate various inexplicable phenomena observed in particle physics experiments. For instance, they could play a significant role in unraveling the mystery surrounding DM \cite{Abazajian:2001nj,Abazajian:2017tcc, Drewes:2016upu}. Sterile neutrinos could also offer a natural explanation for the small active neutrino masses inferred from neutrino oscillation data
\cite{miniboon}.

The idea of right-handed neutrinos, known as sterile neutrinos, is highly plausible as all other fermions have been observed with both left and right chirality, while active neutrinos have only been detected in a left-handed state. Moreover, they could provide a natural explanation for the tiny active neutrino masses deduced from neutrino oscillation experiments\cite{Boyarsky:2018tvu}.
If sterile neutrinos exist, there must be a minimum of three types present to support the ideas of leptogenesis and DM.
This requirement is in contrast to the necessity of having exactly three active neutrino types for ensuring the anomaly cancellation in electroweak interactions   \cite{Ibe:2016yfo}. \par

Here, we briefly describe the ultraviolet (UV) completion of the low-energy effective model adopted in this work. On the one hand, as shown in low-energy experiments, the SM possesses parity-violating (chiral) gauge symmetries $SU_c(3)\times SU_L(2)\times U_Y(1)$. 
On the other hand, as a well-defined quantum field theory, the SM must remain regularized at the high-energy cutoff $\Lambda_{\rm cut}$, fully preserving the SM gauge symmetries.  
A natural UV regularization is provided by a theory of new physics at $\Lambda_{\rm cut}$, for instance, quantum gravity. However, the natural UV regularization and the bilinear fermion Lagrangian of SM chiral gauge symmetries have a theoretical inconsistency due to the No-Go theorem~\cite{NIELSEN1981173, NIELSEN1981219}. This inconsistency suggests the existence of right-handed neutrinos and their quadrilinear four-fermion operators at the UV cutoff scale $\Lambda_{\rm cut}$. Therefore,
we adopt the four-fermion operators of the torsion-free Einstein-Cartan Lagrangian with SM leptons $\psi^{f}$ and three right-handed sterile neutrinos $\nu^{f}_{_R}$~\cite{Xue:2016dpl, Xue:2016txt}:
\begin{eqnarray}
	{\mathcal L}
	&\supset &-G_{\rm cut}\sum_{l=1,2,3}\left(\, \bar\nu^{lc}_{_R}\nu^{l}_{_R}\bar\nu^{l}_{_R} \nu^{lc}_{_R}
	+\, \bar\nu^{lc}_{_R}\psi^{l}_{_R}\bar\psi^{l}_{_R} \nu^{lc}_{_R}\right)+{\rm h.c.},
	\label{art1}
\end{eqnarray}
where the two-component Weyl fermions $\nu^{l}_{_R}$ and $\psi^{l}_{_R}$ are massless chiral fermions. 
	The effective four-fermion operators in Eq.(\ref{art1}) are relevant to the topic studied in this article. We do not consider other types of four-fermion operators in Refs.~\cite{Xue:2016dpl, Xue:2016txt}. In these four-fermion operators, which fully preserve the SM chiral gauge symmetries, left- and right-handed fermions are respectively the eigenstates of the SM chiral gauge symmetries.     
	We assume that the quantum gravity or other new physics that induces four-fermion interactions (\ref{art1}) at the UV cutoff $\Lambda_{\rm cut}$
	should be fermion-flavour blind. Thus, the four-fermion coupling $G_{\rm cut}\propto \Lambda^{-2}_{\rm cut}$ is flavor independent. Thus, the four-fermion operators at the UV cutoff possess exactly the global flavour symmetries $U_L(3)\times U_R(3)$, and one can make unitary transformations to obtain the four-fermion operators like (\ref{art1}) diagonalised in the flavour space. Therefore, non-flavour exchanging interactions (both charged and neutral channels) can occur at the Lagrangian (tree) level of the four-fermion operators (\ref{art1}), detailed discussions can be found in Ref.~\cite{Xue:2016txt}. These operators possess (i) 
	a strong coupling phase where composite particles are formed and symmetries are preserved, (ii) a weak coupling phase where  
	spontaneous symmetry breaking occurs and elementary particles acquire masses. The details are given in Refs.~\cite{Xue:2016dpl, Xue:2016txt} and \cite{Xue:2022kjl}. In this paper, we will consider the second possibility, i.e. the symmetry breaking phase.

In the four-fermion interactions (\ref{art1}), the neutrino self-interaction undergoes spontaneous symmetry breaking to generate the Majorana masses $M^l_{\rm M}=-G_{\rm cut}\langle\bar\nu^{lc}_{_R}\nu^{l}_{_R}\rangle$. \textcolor{black}{It is accompanied by a Goldstone boson (Axion or Majoron) and a massive scalar $\chi$-boson. The properties and observational consequences of these are discussed in Ref.~\cite{Xue:2020cnw}. The neutrino Dirac masses $M^l_{\rm D}$ are generated by usual Yukawa interactions involving the Higgs field, right-handed neutrinos and left-handed leptons. Therefore, neutrino mass terms consist of the Majorana mass $M^\ell_{\rm M}$ and Dirac mass $M^\ell_{\rm D}$},
\begin{eqnarray}
	M^l_{\rm M}\bar\nu^{lc}_{_R}\nu^{l}_{_R} + M^l_{\rm D}\bar\nu^{l}_{_L}\nu^{l}_{_R} +{\rm h.c.}
	\label{masses}
\end{eqnarray}
\textcolor{black}{Here, we suppose that $M^\ell_M$ and $M^\ell_D$ are free parameters and $M^\ell_M\gg M^\ell_D$ \cite{Xue:2016dpl}}.
Through the seesaw mechanism, three Majorana active 
neutrinos ($\nu^{l}_{_L}+\nu^{lc}_{_L}$) masses are given by $M_{\nu_l}\approx (M^l_D)^2/(4M^l_M)$. Three Majorana sterile neutrinos ($\nu^{l}_{_R}+\nu^{lc}_{_R}$) masses are given by $M_{N_l}\approx M^l_M$.  

\begin{figure}
	\centering
	\includegraphics[width=2.2in]{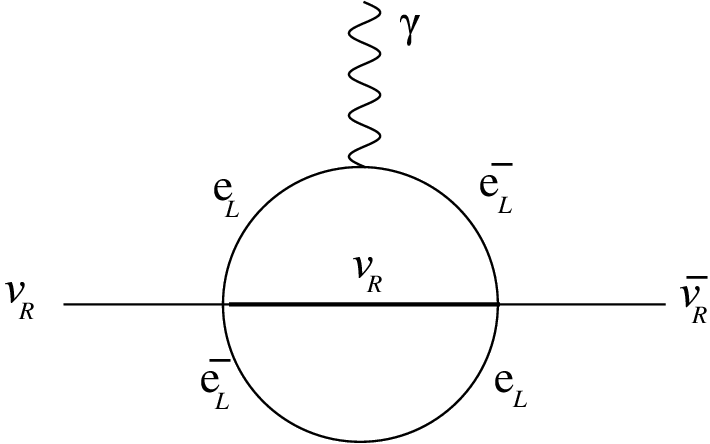}\hskip1.5cm  \includegraphics[width=2.2in]{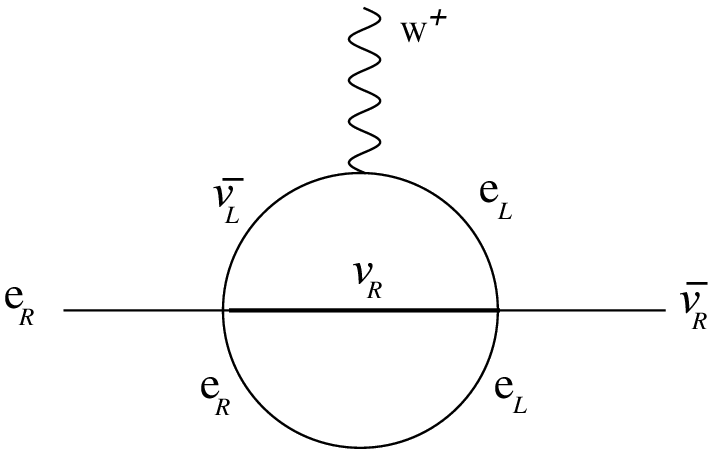}
	\caption{\textcolor{black}{The possible sunset diagrams 
		from the second term of (\ref{art1}), namely, sterile neutrino and SM fermion four-fermion interactions $\bar\nu^{fc}_{_R}\psi^{f}_{_R}\bar\psi^{f}_{_R} \nu^{fc}_{_R}$, in which $\psi^{f}_{_R}$ represents SM right-handed fermions. Hence, these 1PI vertices lead to effective SM gauge boson couplings to right-handed neutrinos (\ref{rhc0}). 
		Left: the effective 
		1PI interacting vertex (\ref{rhc0}) of the gauge boson $W^+$ and right-handed 
		sterile neutrino $\nu^\ell_R$, for more details see Figure 3 of Ref.~\cite{Xue:2015wha}. 
		Right: the effective 1PI interacting vertex (\ref{rhc0}) 
		of photon $\gamma$ and right-handed sterile neutrino $\nu^\ell_R$, and similar one 
		for $Z^0$ boson. The slightly thick solid lines inside sunset diagrams 
		represent right-handed neutrino propagators with Dirac mass 
		(left) or Majorana mass (right). 
		A Dirac mass term is present in the internal electron propagator 
		from $e_L$ to $e_R$ in the left sunset diagram. \cite{Xue:2020cnw}. }}
	\label{Rcouplingf}
\end{figure}
\color{black}

On the other hand, the 
second four-fermion operator in (\ref{art1}) shows
neutrinos' interactions with SM charged leptons $\psi^l_R$. These
effectively {\it{induce}} at low energies the one-particle-irreducible (1PI) vertices \cite{Xue:2020cnw}
\begin{eqnarray}
	\mathcal{L}&\supset & {\mathcal{G}^W_R}~({g_w}/{\sqrt{2}})\bar \ell_R\gamma^\mu\nu^\ell_RW^-_\mu 
	+{\mathcal{G}^Z_R}~({g_w}/{\sqrt{2}})\bar \nu^\ell_R\gamma^\mu\nu^\ell_R Z^0_\mu\nonumber\\ 
	&+& {\mathcal{G}^\gamma_R}~(e)\bar \nu^\ell_R\gamma^\mu\nu^\ell_R A_\mu + {\rm h.c.}\,
	\label{rhc0}
\end{eqnarray}
of right-handed currents interacting with the SM gauge bosons.\par

 It is important to emphasize that the four-fermion interaction of only right-handed sterile neutrinos in the Lagrangian (\ref{art1}) does not induce the effective right-handed coupling of the $W$ boson. Essentially, the four-fermion interactions introduce a small mixing between right-handed neutrinos and left-handed leptons, leading to an effective $W$ boson coupling to right-handed currents. To be more clear, we have provided Fig. \ref{Rcouplingf} in which  the sunset Feynman diagrams depict the one-particle-irreducible (1PI)  interacting vertices, showing that these vertices can be effectively induced from the second four-fermion operator of Eq.~(\ref{art1}).
This reveals that right-handed neutrinos interact with SM leptons. In other words, the four-fermion interactions introduce the small mixing between right-handed neutrinos and left-handed leptons, leading to an effective
 	coupling of W and right-handed current \footnote{The four-fermion interaction of only right-handed sterile neutrinos in the Lagrangian (\ref{art1}) does not induce the effective right-handed coupling of  $W$ boson.}.
 	The effective couplings ${\mathcal{G}^{W,Z,\gamma}_R}$ are energy-dependent functions and are generally small at low energies. In this paper, we treat them as effective parameters. Moreover, it is worth mentioning that the resulting Axion or Majoron in the model has a small coupling to SM particles, which is linked to the smallness of the effective right-handed coupling ${\mathcal{G}_R}$, rather than heavy sterile neutrino masses (see Ref.~\cite{Xue:2020cnw} for detailed discussions and calculations).\par

\begin{figure}[tb]
	\center
	\includegraphics[scale=0.9]{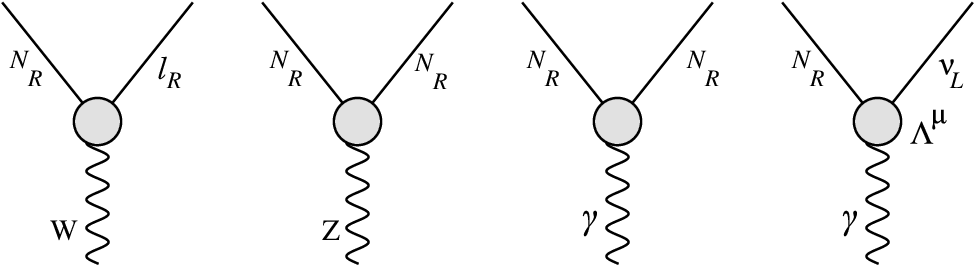}\\
	\caption{The effective coupling of right-handed current with W , Z bosons and photon, \textcolor{black}{which are not present in the SM. The corresponding expressions are Eqs.~(\ref{rhc01}) and (\ref{effem1}).}}. \label{fig1}
\end{figure}

Moreover, we present the Feynman diagrams corresponding to these possible interaction vertices in Fig.~\ref{fig1}. The two diagrams on the left correspond to the interaction of the sterile neutrino with $W$ and $Z$ bosons based on the Eq.~(\ref{rhc0}),  and sterile neutrino -photon vertices in (\ref{rhc0}) are shown in two right diagrams.
The charged current interacts with $W^\pm_{\mu}$ gauge boson and the neutral currents interact with $A^\mu$ photon and $Z_\mu$ boson. The SM gauge couplings $g_w=e/\sin\theta_W$, the electric charge $e$ and Weinberg angle $\theta_W$ relate to 
the Fermi constant as $G_{F}/\sqrt{2}=g_{w}^{2}/8M_{W}^{2}$. 
The effective right-handed couplings ${\mathcal{G}^W_R}$, ${\mathcal{G}^Z_R}$ and ${\mathcal{G}^\gamma_R}$ are small dimensionless parameters beyond the SM. 
Their upper limits must be constrained by Earth-based experiments and astrophysical and cosmological observations. We assume they are of the same order and 
adopt unique notation ${\mathcal{G}_R}\ll 1$. We have not been able to fix the value of \( \mathcal{G}_R(\mu) \), which is expected to be proportional to \( \mathcal{O}\left[(\mu/\Lambda)^2\right] \), where \( \mu \) corresponds to the electroweak (SM) energy scale and \( \Lambda \sim \) TeV denotes the compositeness scale associated with the UV fixed point of a strongly coupled four-fermion interaction~\cite{Xue:2016txt, Leonardi:2018jzn}.

The right-handed neutrino $\nu^\ell_R$ belongs to the same family as the charged lepton $\bar\ell_R$. The right-handed doublets $(\nu^\ell_R, \ell_R)$ are gauge eigenstates, and $(N^\ell_R, \ell_R)$ are the corresponding mass eigenstates. 
In terms of mass eigenstates $( N^l_R,l_R)$, 
gauge eigenstates $\nu^\ell_R = (U^\nu_R)^{\ell l'}N^{l'}_R$ and $\ell_R= (U^\ell_R)^{\ell l'} l'_R$, where $U^{\nu}_R$ and $U^{\ell}_R$ are $3\times 3$ unitary matrices in family flavor space, the 1PI interactions (\ref{rhc0})  take the following form
\begin{eqnarray}
\mathcal{L}&\supset & {\mathcal{G}^W_R}~({g_w}/{\sqrt{2}})[(U^l_R)^\dagger U_R^\nu]\bar \ell_R\gamma^\mu N^\ell_RW^-_\mu 
+{\mathcal{G}^Z_R}~({g_w}/{\sqrt{2}})\bar \nu^\ell_R\gamma^\mu\nu^\ell_R Z^0_\mu\nonumber\\ 
&+& {\mathcal{G}^\gamma_R}~(e)\bar \nu^\ell_R\gamma^\mu\nu^\ell_R A_\mu + {\rm h.c.}\,
\label{rhc01}
\end{eqnarray}
where the flavor mixing matrix $V_R^{ll'}=[(U^l_R)^\dagger U_R^{\nu_{l'}}]$ appears in charged current interaction, which induces the flavour-mixing interactions among different charged sectors and flavours, as those in the SM. However, the $V_R^{ll'}$ are new mixing matrix elements, and differs from the Pontecorvo-Maki-Nakagawa-Sakata (PMNS) one $V_L^{ll'}=[(U^l)^\dagger_L U^{\nu_{l'}}_L]$, which associates to the SM left-handed current interaction $\bar \nu^l_L\gamma^\mu\ell^l_L W_\mu$.   
	While the neutral current interactions in (\ref{rhc01}) remain diagonal in lepton family flavour space, and summation over three lepton families is performed. No flavour-changing-neutral-current (FCNC) interactions occur in Eq.~(\ref{rhc01}). \par 

At the high-energy cutoff $\Lambda_{\rm cut}$, 
	the four-fermion coupling $G_{\rm cut}$ (\ref{art1}) is flavour independent; massless fermions are in the SM gauge eigenstates.  While in the symmetry-breaking ground
	state at low energies, experimentally measured final states of massive fermions mix in different charged sectors and flavours by unitary transformations $U_{L,R}^{l,\nu}$ aforementioned. 
	The recent study on leptoquarks associated with 
	the four-fermion operators at the LHC and HL-LHC has 
	considered flavour mixing across SM families \cite{Ajmal:2023yhl}. The mixing matrix elements of this type should be small, but their effects need more investigation. On the other hand, the effective flavour-exchanging 1PI interaction can be generated by the $W$ boson interacting 
	at the loop level. Therefore, it is worthwhile to study the induced flavour-exchanging process, like B-physics anomalies in LEP and lepton-flavour-changing processes 
	$\mu\rightarrow e + \gamma$ \cite{newemu}. We leave for a future effort the detailed analysis in this context.

The effective operator $\mathcal{G}^{W}_{R}$ contributes to vector boson fusion (VBF) processes, as illustrated by the left Feynman diagram in Fig. 1 of 
Ref.~\cite{CMS:2023nsv}. Moreover, the constraints on 
the sterile neutrinos' mixing $|V_R|$ and masses $M_{N^\ell}$ have been studied \cite{CMS:2023nsv,Sirunyan2018,Sirunyan2019}, indicating that 
the upper bound on the value of ${\mathcal{G}_R}$ must be smaller than $10^{-4}$. This upper limit is further supported by studying the ratio of top- and bottom-quark masses \cite{Xue:2015wha}, the double beta-decay $0\nu\beta\beta$ experiment \cite{Pacioselli:2020yzx}, $W^\pm$ and $Z^0$ decay widths \cite{Haghighat:2019rht}, $W$ 
boson mass tension \cite{Xue2022a}, CMB cosmic birefringence \cite{Mahmoudi:2024wjy}, and 
the precision measurement of fine-structure constant $\alpha$ \cite{Xue:2020cnw}.

 Regarding $N^e_R$ as a DM particle in the XENON1T experiment and astrophysical observations \cite{ Shakeri2020}, one studied the 
relevant 1PI vertex $[(U^\nu_L)^\dagger U^\ell_L]^{l l'} \bar\nu_L^l\Lambda^\mu_{l'} \nu_R^{l'} A_\mu$,
\begin{equation}
	\Lambda_{l'}^\mu(q) =i\frac{e g_w^2 {\mathcal G}_Rm_{l'}}{16\pi^2}\Big[(C_0+2C_1)p_1^\mu+(C_0+2C_2)k_1^\mu\Big]\,
	\label{effem1}
\end{equation}
which is induced by the effective Lagrangian (\ref{rhc0}). \textcolor{black}{Depicted in the last Feynman diagram in Fig.~\ref{fig1}, this 1PI vertex $\Lambda^\mu_{l'}$ belongs to the effective interacting Lagrangian (\ref{rhc01}) of the leading order ${\mathcal G}_R$.} 
Here $p_{1}^{\mu}$ and $k_{1}^{\mu} $ represent the four-momenta of incoming sterile neutrinos and outgoing SM neutrinos, respectively. The three-point Passarino-Veltman functions \cite{PASSARINO1979151} $C_0$, $C_1$ and $C_2$ approach $M_W^{-2}$ in the zero momentum transfer limit $q^{2}=(k_1-p_1)^{2}\rightarrow 0$. Among possible induced 1PI operators in low energies, we study in this article relevant ones possibly accounting for the Muon $a_{\boldsymbol{\mu}}$ anomaly. 

\section{Prediction on the $\mu$AMM }
\label{sec4}
In this section, we aim to calculate the magnetic dipole form factor ($F_2$) for the muon particle while considering the contributions from the right-handed sterile neutrinos using the effective 1PI operators (\ref{rhc01}) and (\ref{effem1}).

\subsection{Contribution due to  W boson and $\nu_{R}$ mediation: Standard neutrino interactions} \label{sec4-1}

To this end, we first need to sketch all feasible one-loop Feynman diagrams involving W boson and $\nu_{R}$ mediation. In our case, there are three novel vertex corrections shown in Fig. (\ref{fig2}), where below, you can find the vertex correction for each of these three diagrams $(a)$, $(b)$ and $(c)$: 
\begin{figure}[tb]
	\center
	\includegraphics[scale=0.95]{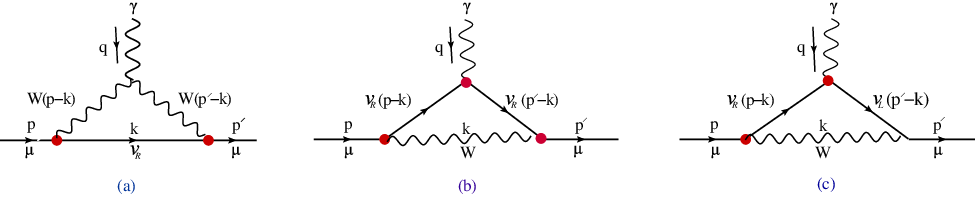}\\
	\caption{The contributions of sterile neutrinos $\nu_R$,  SM (Active) neutrinos $\nu_L$ and  W bosons to the one-loop vertex correction of the muon anomalous magnetic moment. \textcolor{black}{The red spots indicate the effective interacting vertices depicted in Fig.~\ref{fig1}, which are absent in the SM.}} \label{fig2}
\end{figure} 
\begin{eqnarray}
	\label{eq: amp1}
	\bar{u}(p'^{})\left(-{ i} e \Gamma^{\mu}_{(a)}\right)u(p^{})
	&=&{- i} e\,\, V^{{\boldsymbol{\mu}} l}_R\,\,V^{\dagger{\boldsymbol{\mu}} l}_R\int \frac{{\rm d}^4 k}{\left(2\pi\right)^4}\nonumber\\&&\times\left\{\bar{u}(p'^{})(\frac{-i\mathcal{G}_Rg_w}{2\sqrt{2}})\gamma^\nu\left(1+\gamma_5^{}\right)\frac{{ i}}{\slashed{k}-M_{N_{l}}}(\frac{-i\mathcal{G}_Rg_w}{2\sqrt{2}})\gamma^\rho\left(1+\gamma_5^{}\right)\right.\nonumber\\
	&&\left. \times u(p^{})
	\frac{{- i}}{\left(p-k\right)^2-M_W^2}\frac{{- i}}{\left(p'-k\right)^2-M_W^2}\left[g_{\rho \alpha}-\frac{(p-k)_\rho (p-k)_\alpha}{M_W^2}\right]\right.\nonumber\\
	&& \left. \times \left[g_{\nu \beta}-\frac{(p'-k)_\nu (p'-k)_\beta}{M_W^2}\right]\left[g_{}^{\beta\alpha}\left(2k-p^{}-p'^{}\right)_{}^{\mu}\right.\right.\nonumber\\
	&&\left.\left. +g_{}^{\alpha\mu}\left(2p^{}-p'-k\right)_{}^{\beta} +g_{}^{\mu\beta}\left(2p'^{}-k-p^{}\right)_{}^{\alpha}
	\right]
	\right\},
\end{eqnarray}
\begin{eqnarray}
	\label{eq: amp2}
	\bar{u}(p'^{})\left(-{\rm i} e \Gamma^{\mu}_{(b)}\right)u(p^{})&=&{- i} e \,\,V^{{\boldsymbol{\mu}} l}_R\,\,V^{\dagger{\boldsymbol{\mu}} l}_R\int \frac{{\rm d}^4 k}{\left(2\pi\right)^4}\nonumber\\&&\times\left\{\bar{u}(p'^{})(\frac{-i\mathcal{G}_Rg_w}{2\sqrt{2}})\gamma^\nu\left(1+\gamma_5\right)\frac{{i}}{\slashed{p'}-\slashed{k}-M_{N_{l}}}{i}\mathcal{G}_R^\gamma \gamma^\mu\left(1+\gamma_5^{}\right)
	\right.\nonumber\\
	&&\left. \times
	\frac{{ i}}{\slashed{p}-\slashed{k}-M_{N_{l}}}
	(\frac{-i\mathcal{G}_Rg_w}{2\sqrt{2}})\gamma^\rho\left(1+\gamma_5^{}\right) u(p^{})
	\frac{{-i}}{k^2-M_W^2}\left[g_{\rho \nu}-\frac{k_\rho k_\nu}{M_W^2}\right]
	\right\},
\end{eqnarray}
\begin{eqnarray}
	\label{eq: amp3}
	\bar{u}(p')\left(-{\rm i} e \Gamma^{\mu}_{(c)}\right)u(p^{})
	&=&{- i} e \,\,V^{{\boldsymbol{\mu}} l}_R\,\,\int \frac{{\rm d}^4 k}{\left(2\pi\right)^4}\nonumber\\&&\times\left\{\overline{u}(p'^{})(\frac{-{ i} g_{w}}{2\sqrt 2})\gamma^\nu\left(1-\gamma_5^{}\right)\frac{{ i}}{\slashed{p'}-\slashed{k}-M_{N_{l}}}{ i}(\frac{-\mathcal{G}_R g_{ w}^2m_{l}}{16\pi^2})
	\right.\nonumber\\
	&&\left. \times
	\Big((C_0+2C_1)(p'-k)^\mu+(C_0+2C_2)(p-k)^\mu\Big)
	\frac{{ i}}{\slashed{p}-\slashed{k}-M_{\nu_l}}
	\right.\nonumber\\
	&&\left. \times
	(\frac{-i\mathcal{G}_Rg_w}{2\sqrt{2}})\gamma^\rho\left(1+\gamma_5\right) u(p)
	\frac{{-i}}{k^2-M_W^2}\left[g_{\rho \nu}-\frac{k_\rho k_\nu}{M_W^2}\right]
	\right\},
\end{eqnarray}
in which $u(p)$ is the ordinary free Dirac spinor, $M_W$, $M_N$, and $M_\nu$ are the masses of the W boson, sterile neutrino, and SM left-handed neutrino, respectively. Indices $(a)$, $(b)$, and $(c)$ on the left-hand side pertain to the Feynman diagrams depicted in Fig.(\ref{fig2}).
After performing straightforward computations at zero momentum transfer ($q^2\rightarrow 0$) and singling out the part to be proportional to $(p+p')^\mu$, the contribution made by the right handed sterile neutrino into the $\mu$AMM will be obtained as follows
\begin{eqnarray}
	\Delta a^{SN (a)}_{\boldsymbol{\mu}}=2(\frac{\mathcal{G}_Rg_w}{4\pi})^{2}\,m_{\boldsymbol{\mu}}^2V^{\mu l}_R V^{\dagger\mu l}_R\int_{0}^{1} \,dx\,\int_{0}^{1-x}\,dy
	\,\frac{2y(1+3x)+(1-x)^2-8}{(1-x){M_W^2+x M_{N_{l}}}^2-x(1-x)m_{\boldsymbol{\mu}}^2}
	,\label{gamma181}
\end{eqnarray}
\begin{eqnarray}
	\Delta a^{SN(b) }_{\boldsymbol{\mu}} = -(\frac{\mathcal{G}_Rg_w}{\pi})^{2}\mathcal{G}_{R}^{\gamma}m_{\boldsymbol{\mu}}^2V^{\mu l}_R V^{\dagger\mu l}_R\int_{0}^{1} dx\int_{0}^{1-x}\,dy\,\frac{1-2x+y(x+y-2)-(1-x-y)^2}{x M_{W}^2+(1-x)M_{N_l}^2- x (1-x)m_{\boldsymbol{\mu}}^2}
	,\label{gamma182}
\end{eqnarray}
\begin{eqnarray}
	\Delta	a^{SN(c) }_{\boldsymbol{\mu}}=\frac{\mathcal{G}_R^{2}g_w^{4}V^{{\boldsymbol{\mu}} l}_R}{64\pi^4}m_{\boldsymbol{\mu}}m_{l}(C_0+2C_1)\int_{0}^{1} dx\int_{0}^{1-x}dy\frac{m_{\boldsymbol{\mu}}^2(1-2(1-x)+(1-x)^2)+M_{N_l}M_{\nu}}{x M_{W}^2+(1-x-y)M_{N_l}^2+yM_{\nu}^2- x (1-x)m_{\boldsymbol{\mu}}^2}\nonumber\\
	.\label{gamma183}
\end{eqnarray}
where the superscript SN stands for the sterile neutrino effects and the summation $l=e,\boldsymbol{\mu},\tau$ is over three lepton families. 
In the Appendix (\ref{appe}), we present more details of the calculation to achieve the result (\ref{gamma181}). The contributions of the Feynman diagrams (b) and (c) are obtained in a similar way.\par

The total sterile neutrino contribution on $\mu$AMM is given by,
\begin{eqnarray}
	\Delta a_{\boldsymbol{\mu}}^{SN}&=&\Delta a_{\boldsymbol{\mu}}^{SN(a)}+\Delta a_{\boldsymbol{\mu}}^{SN(b)}+\Delta a_{\boldsymbol{\mu}}^{SN(c)}.
	\label{asn}
\end{eqnarray}
However, due to the smallness of the right-handed couplings $\mathcal{G}_R$, $SU_L(2)$ coupling $g_w$ and active neutrino masses $M_\nu$, the main contribution arises from $a_{(a)}^{SN}$. Hence, in the following, the contribution of diagram (a) will be taken account while the contributions of the other diagrams will be ignored. \par

Considering Eq. \ref{gamma181} and performing the integrations, the following result is obtained
\begin{eqnarray}
\Delta a^{SN (a)}_{\boldsymbol{\mu}}&=&2(\frac{\mathcal{G}_Rg_w}{4\pi\,m_{\boldsymbol{\mu}}^2})^{2}\,V^{\mu l}_R V^{\dagger\mu l}_R \nonumber\\
&&\bigg\{	m_{\boldsymbol{\mu}}^4+2 m_{\boldsymbol{\mu}}^2 \left(M_W^2-M_{N_{l}}^2\right)+2\left(3 m_{\boldsymbol{\mu}}^4-m_{\boldsymbol{\mu}}^2 M_{N_{l}}^2+\left(M_W^2-M_{N_{l}}^2\right)^2\right) \ln \left(\frac{M_{N_{l}}}{M_W}\right)\nonumber\\
	&&+\frac{1}{ \sqrt{A}}\Bigg(2 \left[3 m_{\boldsymbol{\mu}}^6+ m_{\boldsymbol{\mu}}^4 \left(4 M_{N_{l}}^2-3 M_W^2\right)+ m_{\boldsymbol{\mu}}^2 \left(M_W^4+M_W^2 M_{N_{l}}^2-2 M_{N_{l}}^4\right)-\left(M_W^2-M_{N_{l}}^2\right)^3\right] \nonumber\\
	&&\left[\tan ^{-1}\left(\frac{- m_{\boldsymbol{\mu}}^2-M_W^2+M_{N_{l}}^2}{\sqrt{A}}\right)-\tan ^{-1}\left(\frac{m_{\boldsymbol{\mu}}^2-M_W^2+ M_{N_{l}}^2}{\sqrt{A}}\right)\right]\Bigg)\bigg\}
	\label{exactsol}
\end{eqnarray}
where  $A=-m_{\boldsymbol{\mu}}^4+2 m_{\boldsymbol{\mu}}^2(M_W^2+M_{N_{l}}^2)-(M_W^2- M_{N_{l}}^2)^2$. \par

Here, we are interested in the situations in which the mass of sterile neutrinos is approximately equal to $M_{W}$. Under this condition, the relation \eqref{exactsol} can be approximated as follows
\begin{eqnarray}
	\Delta a_{\boldsymbol{\mu}}^{SN(a)}&\approx&\frac{19\mathcal{G}_R^{2}g_w^{2}m_{\boldsymbol{\mu}}^2}{3(4\pi)^2M_W^2}
	V^{{\boldsymbol{\mu}} l}_R V^{\dagger{\boldsymbol{\mu}} l}_R.
	\label{resulteach1}
\end{eqnarray}
\begin{figure*}[t!]
	\centering
	\centering
	\includegraphics[width=0.5\linewidth]{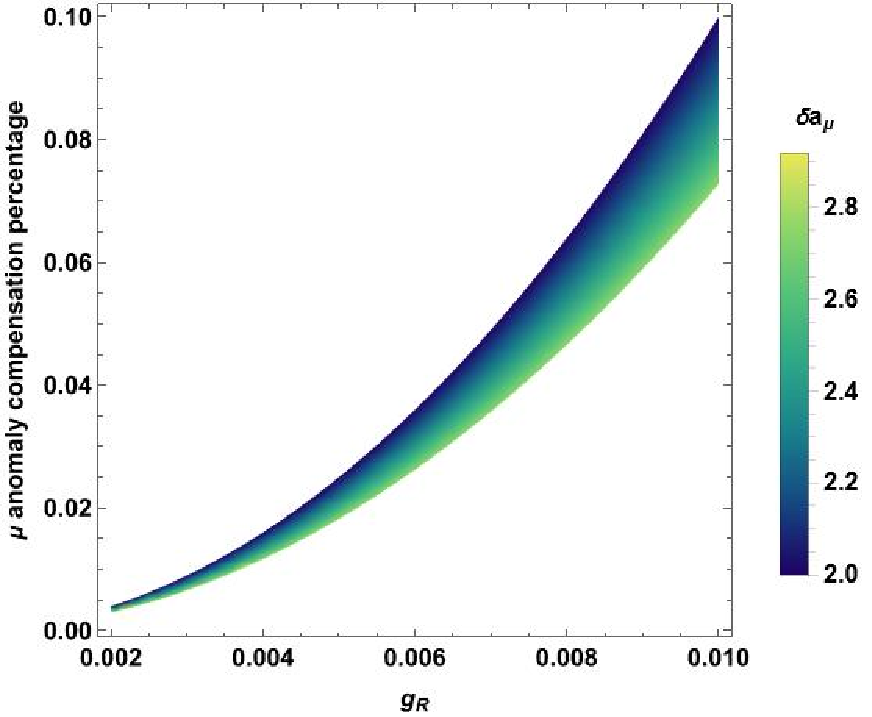}
	\caption{The percentage of the $a_{\boldsymbol{\mu}}$ anomaly compensation in terms of  $\mathcal{G}_R$. Note: $\delta a_{{\boldsymbol{\mu}}}$ stands for $\delta a_{{\boldsymbol{\mu}}}=10^9 \Delta a_{{\boldsymbol{\mu}}}$}
	\label{param2}
\end{figure*}
Based on this relation, one can determine what percentage of the ${\boldsymbol{\mu}}$ anomaly can be compensated by this right-handed sterile neutrino interaction. This is due to the right-handed neutrino and left-handed lepton mixing introduced by the four-fermion interaction in (\ref{art1}). In this regard, we have provided Fig. \ref{param2} where the percentage of the ${\boldsymbol{\mu}}$ anomaly compensation ($\frac{\Delta a _{\mu}^{SN}}{\Delta a _{\mu}}\times 100$) is plotted in terms of  $\mathcal{G}_R$. It should be noted that the density plot is confined between the upper and lower experimental values of $\Delta a_{{\boldsymbol{\mu}}}^{SN}$, ($(2.49 \pm 0.48) \times10^{-9}$), respectively. The values of \( \delta a_\mu \) are presented to illustrate how the $\mu$ anomaly compensation percentage varies over the given parameter range. These plots provide a visual understanding of the parameter regions where the model can effectively account for the observed discrepancy in the muon's anomalous magnetic moment. As can be seen from the figure, smaller values of the coupling constant will contribute less to the compensation of the muon anomaly. For instance, the interaction of right-handed sterile neutrinos with a coupling constant of the order of ${\mathcal G}_{R}\approx 0.004$ can account for less than $0.01\%$ of the total anomaly. \par

Before ending this part, it is worth reviewing some results beyond the well-known $\alpha/2\pi$ term for the $\boldsymbol{\mu}$AMM from QED contribution. In this regard, the one-loop contributions to $\boldsymbol{\mu}$AMM due to the electroweak interactions of the SM were calculated quite a long time ago by Bardeen et al. \cite{smgfactor1}.  
They computed approximately 20 Feynman diagrams at the one-loop level and finally showed that the result is finite. The corresponding outcome was given by \cite{smgfactor2}
\begin{eqnarray}
	\label{eq:a1loop} a_{\boldsymbol{\mu}}^{\rm Weak}=\frac{g_w^2m_{\boldsymbol{\mu}}^2}{64\pi^2M_w^2
	}\Biggl\{
	\frac{10}{3}
	+\frac{4}{3}(v_{\boldsymbol{\mu}}^2-5\, a_{\boldsymbol{\mu}}^2) + {\cal
		O}\biggl(\frac{m_{\boldsymbol{\mu}}^2}{M_Z^2}
	\log\frac{M_Z^2}{m_{\boldsymbol{\mu}}^2}\biggr) + 2\int_{0}^{1}
	dx\frac{x^2(2-x)}{x^2+\frac{M_{\rm H}^2}{m_{\boldsymbol{\mu}}^2}(1-x)} \Biggr\} 
\end{eqnarray}
Where $v_{\boldsymbol{\mu}}$ and $a_{\boldsymbol{\mu}}$ denote the vector and axial-vector couplings of the $Z$ boson to the muon, respectively. Moreover, for a fermion $f$:
\begin{eqnarray}
	\label{eq:couplings} v_{f}=I_f^{(3)}-2Q_{f}\sin^2\theta_W\,, \qquad
	a_{f}=I_f^{(3)}\,.
\end{eqnarray} 
It is important to note that Eq.~(\ref{eq:a1loop}) contains additional terms that are divergent and arise from the anomaly that results when the triangle is multiplied by $k_\mu$. However, this anomaly vanishes and the result becomes finite and gauge invariant if one sums over a complete fermion's generation and considers all of the Feynman diagrams \cite{smgfactor2}. \par

Moreover, Bardeen and Lautrup used the approach of dimensional regularization in \cite{smgfactor1} and obtained the contribution of $W$ boson interacting with left-handed active neutrinos $\nu_L$, being the counterpart of the Feynman diagram (a) of Fig.(\ref{fig2}), as follows
\begin{eqnarray}
	\label{eq:muon}
	a_{\boldsymbol{\mu}}^\nu \simeq \,-\frac{1}{64\pi^2}\frac{g_w^2 m^{2}_{\boldsymbol{\mu}}}{M_w^2}\frac{10}{3}
	= - 9.11\times 10^{-11},
	\label{Bardeen}
\end{eqnarray}
where $m_{\boldsymbol{\mu}}$ is the muon mass and the active neutrino mass is set to zero. 
We followed the same approach to calculate the contribution of Fig.(\ref{fig2}a)  for $W$ boson interacting with the right-handed sterile neutrinos $\nu_R^\ell$ in three generations. Compared with the corresponding calculations in SM, the differences come from the right-handed coupling ${\mathcal G}_{R} g_w$, sterile neutrino masses  $M_{N_l}\gg M_{\nu_l}$ and right-handed family mixing matrix $V^{\mu l}_R$ instead of PMNS mixing matrix $V^{\mu l}_L$. In addition, as expected,  $a_{(a)}^{SN}$ reduces to $a_{\boldsymbol{\mu}}^\nu$ (\ref{Bardeen}) when $\mathcal{G}_R^{2}\rightarrow 1$, $M_{N_l}\rightarrow 0$ and $\sum_l V^{\mu l}_R V^{\dagger\mu l}_R=1$.\par


\subsection{Contribution due to W boson mediation with Left-Right neutrino mixing: Non-standard neutrino interactions} \label{sec4-2}

\begin{figure*}[t!]
	\centering
	\centering
	\includegraphics[width=0.4\linewidth]{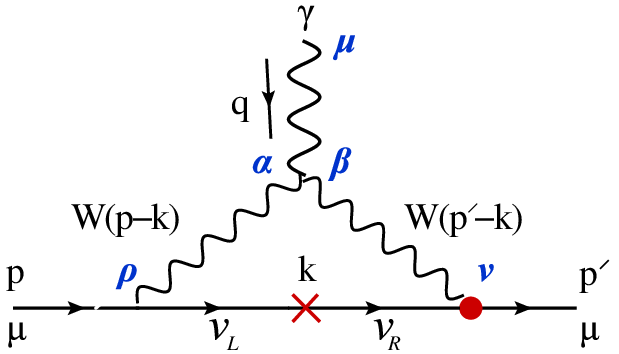}
	\caption{Feynman diagram related to the helicity flip of neutrino mediator. The red cross \textcolor{red}{$({\times})$} represents 
    non-standard neutrino interactions, i.e.,  the Dirac mass insertion. }
	\label{Wein}
\end{figure*}

In this part, we consider another possibility of the right handed sterile neutrino contribution to $\boldsymbol{\mu}$AMM which has been depicted in Fig.\ref{Wein}. Indeed, due to the neutrino mass, a helicity flip may occur in the mediator part, producing an exotic coupling often referred to as {\it{non-standard neutrino interactions }}. This process is 
similar to those that happen for the Weinberg operator, introduced to explain $0\nu\beta\beta$ \cite{PhysRevLett.43.1566}. 
Based on this method, we treat the mass terms in
Eq. \eqref{masses} as an effective 'two-point vertex'. By considering the neutrino as a Dirac particle, $M_{D}$ is the Dirac neutrino mass matrix connecting one massless left-handed neutrino of momentum $k$ 
with the right-handed one of the same momentum. Analogous to the Weinberg operator studied in Fig. 1 and Eq. 6 of Ref. \cite{Fuks:2020zbm}, its graph simplifies to:
\begin{center}
	\includegraphics[width=0.30\linewidth]{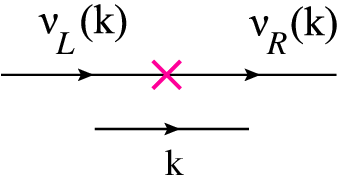}
\end{center}
By assigning the left-handed neutrino coupling vertex $g_w P_L \gamma^\nu$ and the right-handed neutrino coupling vertex ${\mathcal G}_R g_w P_L \gamma^\rho$ to the $W$ gauge boson interactions, we have
	\begin{align}
	{\mathcal G}_Rg_W^2(V_L^{\mu l})^\dagger\gamma^\nu P_R \frac{ i M^l_{D}}{k^2-(M^l_D)^2}P_R \gamma^\rho V_R^{\mu l}
	\label{leftright1}
	\end{align}
 where $l$ family flavors are summed.

From this perspective and making use of \eqref{masses} and \eqref{rhc0}, the vertex correction due to this diagram will be obtained as follows
\begin{eqnarray}
\label{eq: amp11}
\overline{u}(p'^{})\left(-{ i} e \Gamma^{\mu}\right)u(p^{})
&=&{- i} e\, V^{{\boldsymbol{\mu}} l}_L\,V^{\dagger{\boldsymbol{\mu}} l}_R\int \frac{{\rm d}^4 k}{\left(2\pi\right)^4}\left\{\bar{u}(p'^{})(\frac{-i\mathcal{G}_Rg_w}{2\sqrt{2}})\gamma^\nu\left(1+\gamma_5^{}\right)\frac{{ iM_{D}^{l}}}{k^2-(M_{D}^{l})^2}(\frac{-ig_w}{2\sqrt{2}})\gamma^\rho\left(1-\gamma_5^{}\right)\right.\nonumber\\
&&\left. \times u(p^{})
\frac{{- i}}{\left(p-k\right)^2-M_W^2}\frac{{- i}}{\left(p'-k\right)^2-M_W^2}\left[g_{\rho \alpha}-\frac{(p-k)_\rho (p-k)_\alpha}{M_W^2}\right]\right.\nonumber\\
&& \left. \times \left[g_{\nu \beta}-\frac{(p'-k)_\nu (p'-k)_\beta}{M_W^2}\right]\left[g_{}^{\beta\alpha}\left(2k-p^{}-p'^{}\right)_{}^{\mu}\right.\right.\nonumber\\
&&\left.\left. +g_{}^{\alpha\mu}\left(2p^{}-p'-k\right)_{}^{\beta} +g_{}^{\mu\beta}\left(2p'^{}-k-p^{}\right)_{}^{\alpha}
\right]
\right\}.
\end{eqnarray}
Then, by following the steps outlined in Appendix \ref{appe2}, we get the following correction to $a_\mu$ arising from the mentioned diagram of Fig.~(\ref{Wein})
\begin{eqnarray}
	\Delta a^{SN(\times) }_{\boldsymbol{\mu}}&=& -\frac{6g_w^{2}}{(16\pi)^2 m_{\boldsymbol{\mu}}^3}M_{D}^{l}\,\mathcal{G}_R V^{{\boldsymbol{\mu}} l}_L\,V^{\dagger{\boldsymbol{\mu}} l}_R \bigg\{
	-(m_{\boldsymbol{\mu}}^2-M_W^2+M_D^2)\ln{\frac{M_D^2}{M_W^2}}+2m_{\boldsymbol{\mu}}^2
	\nonumber\\&+&\frac{4m_{\boldsymbol{\mu}}^2M_D^2-2A}{\sqrt{A}}(\tan^{-1}(\frac{m_{\boldsymbol{\mu}}^2-M_W^2+M_D^2}{A})-\tan^{-1}(\frac{-m_{\boldsymbol{\mu}}^2-M_W^2+M_D^2}{A}))
	\bigg\}
	.\label{agb1}
\end{eqnarray}
where  $A=-m_{\boldsymbol{\mu}}^4+2m_{\boldsymbol{\mu}}^2(M_W^2+M_D^2)-(M_W^2-M_D^2)^2$. In the limit where $M_{D} \approx M_{W}$, the above relation can be approximated as follows
\begin{eqnarray}
\Delta a^{SN(\times) }_{\boldsymbol{\mu}}&\approx& \textcolor{blue}{-}\frac{2 \mathcal{G}_R g_w^{2}}{(16\pi)^2}\frac{M_{D}^{l} m_{\boldsymbol{\mu}}}{M_{W}^2}\,V^{\mu l}_L V^{\dagger\mu l}_R \,\nonumber\\
&=& \textcolor{blue}{-}\frac{\mathcal{G}_R}{16\sqrt{2}\pi^2}G_{F}\,M_{D}^{l}m_{\boldsymbol{\mu}}\,V^{{\boldsymbol{\mu}} l}_L\,V^{\dagger{\boldsymbol{\mu}} l}_R 
.\label{weinberg2}
\end{eqnarray}
Now, we can use this result to set constraints on $\mathcal{G}_R$. By rewriting the above relation as below
\begin{eqnarray}
\Delta a^{SN(\times) }_{\boldsymbol{\mu}}\approx 4\times 10^{-7} \mathcal{G}_R
 \,\,\,\,\,\,(\frac{m_{\boldsymbol{\mu}}}{105 MeV})\,(\frac{M_{D}^{l}}{100 GeV})\, (\frac{G_{F}}{1.05\times10^{-5} GeV^{-2}})\label{weinberg3}
\end{eqnarray}
and employing the fact that $\Delta a_{\boldsymbol{\mu}}= (2.49 \pm 0.48) \times10^{-9}$, we get the result that $\mathcal{G}_R<10^{-3}$ for $M_{D}^{l}\approx 100 \text{GeV}$. It should be noted that, for the purpose of estimating Eq.~(\ref{weinberg2}), and based on phenomenological arguments~\cite{Wang:2021uqz, Thomson}, the product of flavor mixing matrix elements can be approximated as$
	V^{\boldsymbol{\mu} l}_L V^{\dagger\, \boldsymbol{\mu} l}_R \approx -1.
	$



\begin{figure}
	\centering
	{\includegraphics[scale=0.6]{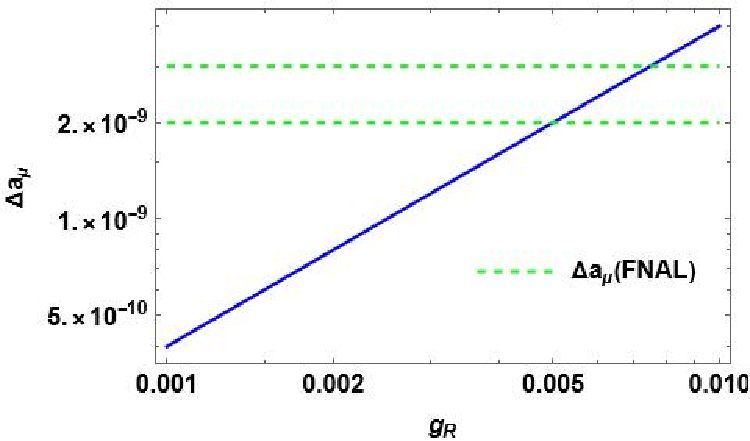}\label{fig5}} \hspace*{2cm} 
	{\includegraphics[scale=0.4]{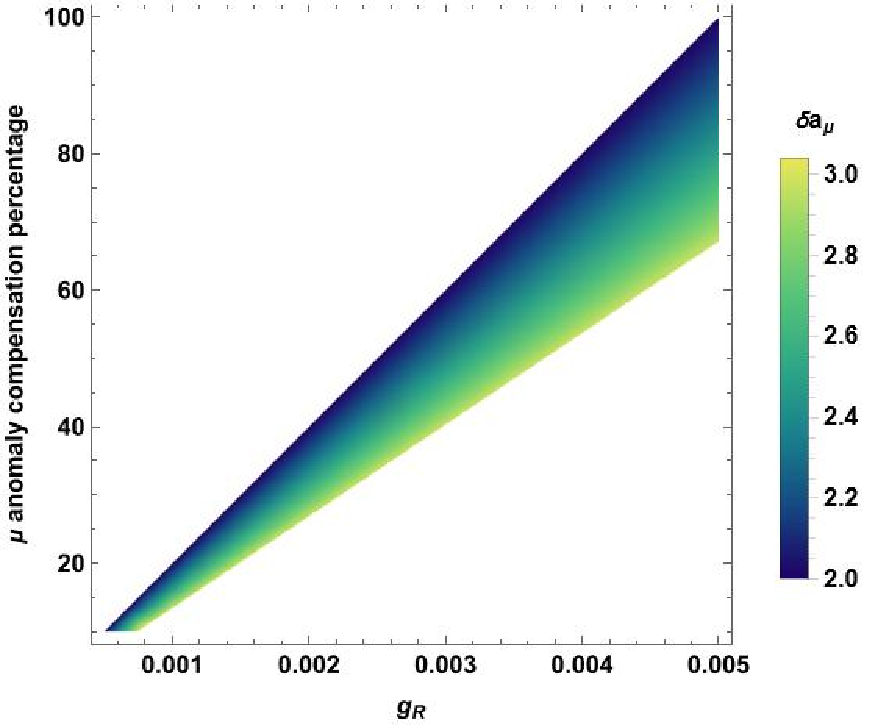}}\label{fig6}
	\caption{Left: The contribution due to the right-handed sterile neutrino interaction with SM particle to $\Delta a_{\mu}$ versus $\mathcal{G}_R$ for $M_{D}^{l}\approx M_{W}$. Right: The percentage of the ${\boldsymbol{\mu}}$ anomaly compensation in terms of $\mathcal{G}_R$. Note: $\delta a_{{\boldsymbol{\mu}}}$ stands for $\delta a_{{\boldsymbol{\mu}}}=10^9 \Delta a_{{\boldsymbol{\mu}}}$. In addition, the density plot is confined between the upper and lower experimental values of $\Delta a_{{\boldsymbol{\mu}}}^{SN}$ ($(2.49 \pm 0.48) \times10^{-9}$), respectively. }
	\label{Wein1}
\end{figure}

The contribution arising from this sort of right-handed sterile neutrino interaction with SM particle to $\Delta a_{\mu}$, for $M_{D}^{l}\approx 100 \text{GeV}$, is presented in the left panel of Fig. \ref{Wein1}. As shown in the figure, for the $\mathcal{G}_R$ coupling constant around 0.005, this type of right-handed neutrino interaction can address the whole anomaly and play a significant role in the context of $\mu$AMM. Moreover, to clarify the impact, we have provided the percentage of the ${\boldsymbol{\mu}}$ anomaly compensation in terms of  $\mathcal{G}_R$ in the right panel of Fig. \ref{Wein1}.\par  

Before concluding this part, it is necessary to mention that one of the most famous possibilities to explain the smallness of neutrino masses is the type-I seesaw mechanism through which SM neutrinos acquire tiny Majorana masses.
	However, the type-I seesaw model suffers from a lack of
	testability, because right-handed neutrinos are too heavy
	to be produced in current collider experiments (For a Dirac mass $M_D^{l}\sim 100$ GeV, it requires a Majorana mass $M_M\gtrsim 10^{12}$ GeV to achieve SM neutrino masses $\lesssim 1 $ eV). To overcome this problem, several models have been proposed that allow for sub-eV neutrino masses while keeping heavy Majorana neutrino masses as low as several hundred GeV \cite{Pilaftsis:1991ug, Kersten:2007vk, Mohapatra:1986bd}. Particularly, the authors in \cite{Zhang:2009ac} have shown that implementing an additional U(1) symmetry and then its soft breaking can veer into having Majorana masses at the order of TeVs. These issues will be topics for our future investigations.

\section{Summery and remarks }
\label{secsumm}

Recent experimental measurements on ${\boldsymbol{\mu}}$AMM exhibit a considerable discrepancy compared to the SM predictions. These differences might be understood in scenarios
of physics beyond SM, involving new particles and interactions. In this work, we investigated the potential contribution of right-handed sterile neutrinos to the $(g-2)_{{\boldsymbol{\mu}}}$ anomaly within an effective four-fermion interacting model based on the fundamental symmetries and particle content of the SM. To this end, we first calculated the magnetic dipole form factor ($F_2$) for the muon particle by considering the standard sterile neutrino interaction which is the interaction including $W$ boson and $\nu_{R}$ mediation. Making use of the effective 1PI operators (\ref{rhc0}) and (\ref{effem1}) and obtaining the correction effects on the ${\boldsymbol{\mu}}$AMM due to the interaction between the right-handed sterile neutrino and the SM particles through the Feynman diagrams depicted in Fig \eqref{fig2}, we found that these kinds of standard neutrino interactions cannot account for the \textcolor{black}{muon $g-2$ anomaly, because as demonstrated in \eqref{resulteach1}, the corrections scale as $m_{\boldsymbol{\mu}}^2/M_W^2$, similar to SM corrections. This should be avoided by a mechanism of chiral enhancement \cite{Crivellin:2021rbq, Crivellin:2018qmi}. Indeed, we obtain the promising correction \eqref{weinberg2}, which is proportional to $m_{\boldsymbol{\mu}} M_D^{l}/M_W^2$,  originating from the chirality flipping a large Dirac mass $M_D^{l}$ of sterile neutrinos}. \par

Next, we considered the non-standard Dirac neutrino interactions being due to $W$ boson mediation with Left-Right neutrino mixing, illustrated via the Feynman diagram in Fig \eqref{Wein}. As our studies showed, if neutrinos can experience such an exotic coupling, they will play an important role in explaining $(g-2)_{{\boldsymbol{\mu}}}$ anomaly.   
In continuation, we tried to analyze the implications of our results in the ${\boldsymbol{\mu}}$AMM and to study the constraint on the parameter space of right-handed sterile neutrino using the experimental data of $a_{\boldsymbol{\mu}}$, which the discussions are as follows:

\begin{itemize}
	\item {
		Making use of the obtained correction term on the $a_{\boldsymbol{\mu}}$, 
		i.e. Eq.~\eqref{weinberg3}, we found that a Dirac mass scale $M_D^{l} \sim 100\text{GeV}$ could explain the muon anomaly if the right handed sterile neutrino's coupling with SM particles is about $\mathcal{G}_R\approx 10^{-3}$. 
}
	
		\item
	{Using the obtained correction term on the $a_{\boldsymbol{\mu}}$ , i.e. Eq.\eqref{weinberg2},, we provided Fig.~\eqref{Wein1} (left panel) for presenting the contribution to $\Delta a_{\mu}$ due to this sort of right-handed sterile neutrino interaction with SM particle. In the right panel of Fig. \eqref{Wein1}, the percentage of the ${\boldsymbol{\mu}}$ anomaly compensation in terms of  $\mathcal{G}_R$ has been plotted. According to this figure, for the $\mathcal{G}_R \approx0.005$ and $M_D^l\approx 100 \text{GeV}$, the right-handed neutrino can address the whole anomaly and play a significant role in the context of $\mu$AMM.}
	\end{itemize}
To further constrain sterile neutrinos' effective mass scale and coupling to SM particles, we necessarily require more experiments and observations.

\textcolor{black}{Before ending this paper it should be noted that the impacts of the effective operators and Feynman diagrams on processes involving different lepton flavors, and explore the constraints by using lepton 
	flavor-violating observables will be investigated in an independent work \cite{newemu}.}\par

Before ending this paper, we would like to note that in the context of this model, the analogous studies can be conducted and the result (\ref{weinberg2}) can be generalised to
	the electron and tau lepton cases 
	\begin{eqnarray}
	\Delta a^{SN(\times) }_{\boldsymbol{\ell}}&\approx&\frac{2 \mathcal{G}_R g_w^{2}}{(16\pi)^2}\frac{M_{D}^{l} m_{\boldsymbol{\ell}}}{M_{W}^2}\,V^{\ell l}_L V^{\dagger\ell l}_R,\quad \ell =e,\tau .\label{etau}
	\end{eqnarray}
	which gives an insight into the electron and muon anomalous magnetic moments $\delta a_{e,\tau}$. The electron (tau lepton) mass $m_e$ ($m_\tau$) is much smaller (larger) than the muon mass $m_\mu$. 
	However the factor $\sum_l M^l_D \,V^{\ell l}_L V^{\dagger\ell l}_R$ remains undetermined.
	The impacts of the effective operators and Feynman diagrams on processes involving different lepton flavours, and exploring the constraints by using lepton flavour-violating observables, will be investigated in separate works.

\section*{ Acknowledgment}

I.~Motie would like to express his gratitude to Professor A.~Blanchard for his gracious hospitality during my stay in Toulouse, France. S.~S.~Xue thanks Drs S.~Shakeri and F.~Hajkarim for discussions.  S. Mahmoudi is grateful to the Iran Science Elites Federation for
the financial support, which contributed to part of the results presented in this paper.

\appendix
\section{Standard Neutrino Interaction Computations}\label{appe}
   
In order to evaluate loop integrals that come from Feynman diagrams, we need to use the Feynman parametrization technique together with the Dirac equation and the standard contraction identity of gamma matrices. For a more in-depth understanding of this method, please see \cite{Peskin:1995ev}.

In the context of Quantum Field Theory (QFT), the computation of vertex corrections of photon-fermion has a unique procedure that can be found in standard QFT textbooks (e.g. see the chapter 6 of \cite{Peskin:1995ev}).\par
In loop integrals, we often encounter products of many propagator factors. To simplify the process of four-momentum integration, we can combine these propagators into a single fraction. This is commonly done using what are known as Feynman parameters \cite{Peskin:1995ev}.
\begin{figure}[tb]
	\center
	\includegraphics[scale=0.45]{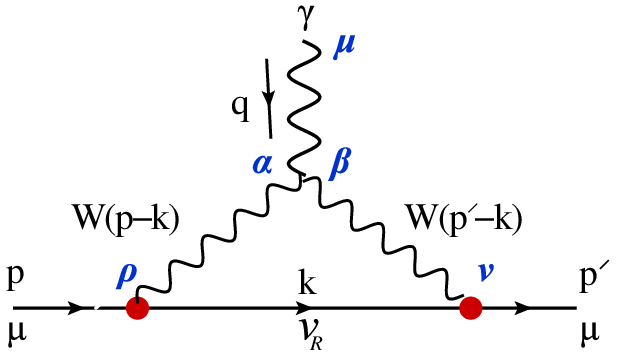}\\
	\caption{One of the Feynman
diagrams which gives
the W-boson contribution to the muon
anomalous moment.} \label{fig3A}
\end{figure} 
\\ 
Here, we provide additional details for Feynman diagram (\ref{fig2}-$a$). For other diagrams, everything remains the same. Considering Eq. \eqref{rhc0} and Fig. \eqref{fig3A}, the vertex correction will be as follows
\begin{eqnarray}
	\label{eqA: amp1}
	\overline{u}(p'^{})\left(-{ i} e \Gamma^{\mu}_{(a)}\right)u(p^{})
	&=&{- i} e\,\, V^{{\boldsymbol{\mu}} l}_R\,\,V^{\dagger{\boldsymbol{\mu}} l}_R\int \frac{{\rm d}^4 k}{\left(2\pi\right)^4}\nonumber\\&&\times\left\{\overline{u}(p'^{})(\frac{-i\mathcal{G}_Rg_w}{\sqrt{2}})\gamma^\nu\left(1+\gamma_5^{}\right)\frac{{ i}}{\slashed{k}-M_{N_{l}}}(\frac{-i\mathcal{G}_Rg_w}{\sqrt{2}})\gamma^\rho\left(1+\gamma_5^{}\right)\right.\nonumber\\
	&&\left. \times u(p^{})
	\frac{{- i}}{\left(p-k\right)^2-M_W^2}\frac{{- i}}{\left(p'-k\right)^2-M_W^2}\left[g_{\rho \alpha}-\frac{(p-k)_\rho (p-k)_\alpha}{M_W^2}\right]\right.\nonumber\\
	&& \left. \times \left[g_{\nu \beta}-\frac{(p'-k)_\nu (p'-k)_\beta}{M_W^2}\right]\left[g_{}^{\beta\alpha}\left(2k-p^{}-p'^{}\right)_{}^{\mu}\right.\right.\nonumber\\
	&&\left.\left. +g_{}^{\alpha\mu}\left(2p^{}-p'-k\right)_{}^{\beta} +g_{}^{\mu\beta}\left(2p'^{}-k-p^{}\right)_{}^{\alpha}
	\right]
	\right\}.
\end{eqnarray}
To evaluate the above integral, we first use the Feynman parameters' method \cite{Peskin:1995ev} to squeeze the three denominator factors into a single quadratic polynomial, raised to the third power, as follows 
\begin{equation}
	\frac{1}{[k^2-M_{N_l}^2+i\epsilon]\left[(p'-k\right)^2-M_W^2+i\epsilon][\left(p-k\right)^2-M_W^2+i\epsilon]}=\int_{0}^{1} dxdydz\delta(x+y+z-1)\frac{2}{D^3}\label{fp}
\end{equation}
where the variable $x$, $y$ and $z$ are called Feynman parameters and the new variable $D$ is defined is given by
\begin{eqnarray}
	D&=&x[k^2-M_{N_l}^2]+y\left[(p'-k\right)^2-M_W^2]+z[\left(p-k\right)^2-M_W^2]\nonumber\\
	&=&k^2-2k.(yp'+zp)-M_w^2(y+z)-x\,M_{N_l}^2+m^2(y+z)+i\epsilon,
	\label{d}
\end{eqnarray}
where in the second line we have used $x+y+z=1$. Now, one can shift $k$ to complete the square
\begin{eqnarray}
	l\equiv k-(yp'+zp),
\end{eqnarray}
and after a bit of calculation, we find that $D$ simplifies to 
\begin{eqnarray}
	D=l^2-\Delta_{(a)}+i\epsilon,
	\end{eqnarray}
	where
	\begin{equation}
		\Delta_{(a)} = -x(1-x)m_{\boldsymbol{\mu}}^2+(1-x)M_W^2+xM_{N_l}^2-zyq^2.
	\end{equation}
In the next step, we must express the numerator of \eqref{eqA: amp1}, i.e. 
\begin{eqnarray}
Numerator &=&
\bar{u}(p')\gamma^\nu\left(1+\gamma_5\right)(\slashed{k}+M_{N_{l}})\gamma^\rho\left(1+\gamma_5\right)\left[g_{\rho \alpha}-\frac{(p-k)_\rho (p-k)_\alpha}{M_W^2}\right]\nonumber\\
&\times& \left[g_{\nu \beta}-\frac{(p'-k)_\nu (p'-k)_\beta}{M_W^2}\right][g^{\beta\alpha}\left(2k-p^{}-p'^{}\right)^{\mu}\nonumber\\
&& +g_{}^{\alpha\mu}\left(2p^{}-p'-k\right)_{}^{\beta} +g_{}^{\mu\beta}\left(2p'^{}-k-p\right)_{}^{\alpha}] u(p),
\end{eqnarray}
in terms of $l$. Making use of the following identities
\begin{eqnarray}
\int\frac{d^{4}l}{(2\pi)^4}\frac{l^\mu}{D^3}&=& 0,\nonumber\\
\int\frac{d^{4}l}{(2\pi)^4}\frac{l^\mu l\nu}{D^3}&=& \int\frac{d^{4}l}{(2\pi)^4}\frac{g^{\mu\nu}l^2}{4\, D^3}     
\label{app2}
\end{eqnarray}
and considering the terms with the highest order, the numerator changes into
\begin{eqnarray}
 Numerator &\rightarrow&\bar{u}(p')\bigg\{\gamma^{\mu}(1+\gamma_5)\Bigg[-3l^2+\Big(z(1-2z)+y(1-2y)+2z(2-y)+2y(2-z)\Big)m_{\boldsymbol{\mu}}^2\nonumber\\
&&-\Big(y(2-z)-z(2-y)\Big)q^2\Bigg]
 (1+\gamma_5)\Bigg[(-2y-2z-4yz)m_{\boldsymbol{\mu}}p^{\mu}-2y(2y-1)
m_{\boldsymbol{\mu}}p'^{\mu}\Bigg]\nonumber\\
&+&(1-\gamma_5)\Bigg[(-2y-2z-4yz)m_{\boldsymbol{\mu}}p'^{\mu}-2z(2z-1)
m_{\boldsymbol{\mu}}p^{\mu}\Bigg]\nonumber\\
&+&
\gamma^{\mu}(1-\gamma_5)\Big(-2zy+y+3z\Big)m_{\boldsymbol{\mu}}^2
\bigg\}u(p)
\end{eqnarray}
Henceforth, we just interested in the terms contributing to the $\mu$AMM. By employing \eqref{form-factor} as well as the Gordon Identity 
\begin{eqnarray}
\bar{u}(p')\gamma^{\mu}u(p)=\bar{u}(p')\bigg (\frac{(p+p')}{2m}^{\mu}+i\frac{\sigma^{\mu\nu}q_{\nu}}{2m}\bigg)u(p)
,\label{Gordon}
\end{eqnarray}
one can find that the terms that contribute to the magnetic moment are those that are multiplied by $\frac{(p+p')}{2m}^{\mu}$. Therefore, the correction on the $\mu$AMM will be obtained as follows
\begin{eqnarray}
	\delta\Gamma^{\mu }_{(a)}&=&-i\,({\mathcal{G}_Rg_w})^{2}V^{\mu l}_R V^{\dagger\mu l}_R\int\frac{d^{4}l}{(2\pi)^4}\int_{0}^{1} \,dx\,dy\,dz\,\delta(x+y+z-1)\frac{2}{(l^{2}-\Delta_{(a)})^{3}}\nonumber\\
	&&\Bigg((-2y-4yz-4z^2)mp^{\mu}+(-2z-4yz-4y^2)mp'^{\mu}\Bigg)
	.\label{gamma16}
\end{eqnarray}
To go further and evaluate the above integral, we use the following relation
\begin{eqnarray}
	\int\frac{d^{4}l}{(2\pi)^4}\frac{1}{(l^2-\Delta)^n}=\frac{(-1)^ni}{(4\pi)^{2}}\frac{\Gamma(n-2)}{\Gamma(n)}(\frac{1}{\Delta})^{n-2},\label{app1}
\end{eqnarray}
and we get
\begin{eqnarray}
\delta\Gamma^{\mu }_{(a)}=-(\frac{\mathcal{G}_Rg_w}{4\pi})^{2}V^{\mu l}_R V^{\dagger\mu l}_R\int_{0}^{1} \,dx\,dy\,dz\,\delta(x+y+z-1)\frac{1}{\Delta_{(a)}}
(-2y-4yz-4z^2)\big(p^{\mu}+p'^{\mu}\big)m
.\label{gamma17}
\end{eqnarray}
Using the above result, we conclude that the correction on the $\mu$AMM due to the participation of the Sterile Neutrino at the zero momentum transfer is as follows
\begin{eqnarray}
a^{SN }_{(a)}&=&-2(\frac{\mathcal{G}_Rg_w}{4\pi})^{2}\,m_{\boldsymbol{\mu}}^2V^{\mu l}_R V^{\dagger\mu l}_R\int_{0}^{1} \,dx\,\int_{0}^{1-x}\,dy\nonumber\\
&\times&\,\frac{2y(1+3x)+(1-x)^2-8}{(1-x){M_W^2+x M_{N_{l}}}^2-x(1-x)m_{\boldsymbol{\mu}}^2}
.\label{gamma18}
\end{eqnarray}

Note that we adopted the $\gamma_5$ anti-commutes with all $\gamma$ matrices. 
Eqs.~(\ref{eq: amp1}), (\ref{eq: amp2}) and (\ref{eq: amp3}) contain a common part, which is divergent and arises from the anomaly that results when the triangle is multiplied by $k_\mu$. The anomaly vanishes and the result becomes finite and gauge invariant only when one sums over a complete generation, 
	as discussed in Ref.~\cite{smgfactor1, smgfactor2}.


\section{Non-Standard Neutrino Interaction Computations}\label{appe2}

Here, we provide additional details for Feynman diagram (\ref{Wein}). Considering Eq. \eqref{eq: amp11}, the vertex correction will be as follows
\begin{eqnarray}
\label{B1}
\overline{u}(p'^{})\left(-{ i} e \Gamma^{\mu}\right)u(p^{})
&=&{- i} e\, V^{{\boldsymbol{\mu}} l}_L\,V^{\dagger{\boldsymbol{\mu}} l}_R\int \frac{{\rm d}^4 k}{\left(2\pi\right)^4}\left\{\bar{u}(p'^{})(\frac{-i\mathcal{G}_Rg_w}{2\sqrt{2}})\gamma^\nu\left(1+\gamma_5^{}\right)\frac{{ iM_{D}^{l}}}{k^2-{M^l_D}^2}(\frac{-ig_w}{2\sqrt{2}})\gamma^\rho\left(1-\gamma_5^{}\right)\right.\nonumber\\
&&\left. \times u(p^{})
\frac{{- i}}{\left(p-k\right)^2-M_W^2}\frac{{- i}}{\left(p'-k\right)^2-M_W^2}\left[g_{\rho \alpha}-\frac{(p-k)_\rho (p-k)_\alpha}{M_W^2}\right]\right.\nonumber\\
&& \left. \times \left[g_{\nu \beta}-\frac{(p'-k)_\nu (p'-k)_\beta}{M_W^2}\right]\left[g_{}^{\beta\alpha}\left(2k-p^{}-p'^{}\right)_{}^{\mu}\right.\right.\nonumber\\
&&\left.\left. +g_{}^{\alpha\mu}\left(2p^{}-p'-k\right)_{}^{\beta} +g_{}^{\mu\beta}\left(2p'^{}-k-p^{}\right)_{}^{\alpha}
\right]
\right\}.
\end{eqnarray}
Agian we need to  first use the Feynman parameters' method \cite{Peskin:1995ev} to squeeze the three denominator factors into a single quadratic polynomial, raised to the third power, as follows 
\begin{equation}
	\frac{1}{[k^2-{M_{D}^l}^2+i\epsilon]\left[(p-k\right)^2-M_W^2+i\epsilon][\left(p'-k\right)^2-M_W^2+i\epsilon]}=\int_{0}^{1} dxdydz\delta(x+y+z-1)\frac{2}{D^3}\label{fp}
\end{equation}
where 
\begin{eqnarray}
	D&=&x[k^2-{M_{D}^l}^2]+y\left[(p'-k\right)^2-M_W^2]+z[\left(p-k\right)^2-M_W^2]\nonumber\\
	&=&k^2-2k.(yp'+zp)-M_w^2(y+z)-x\,{M_{D}^l}^2+m_{\boldsymbol{\mu}}^2(y+z)+i\epsilon.
	\label{d}
\end{eqnarray}
Next, we shift $k$ to complete the square $l\equiv k-(yp'+zp)$. After some calculations, we find that $D$ simplifies to 
\begin{eqnarray}
	D&=&l^2-\Delta+i\epsilon,\nonumber\\
 \Delta &=& -x(1-x)m_{\boldsymbol{\mu}}^2+(1-x){M_W^2+xM_{D}^l}^2-zyq^2.
\end{eqnarray}
The numerator of \ref{B1} is given as 
\begin{eqnarray}
Numerator &=&
\bar{u}(p')\gamma^\nu\left(1+\gamma_5\right)(\slashed{k}+M_{D}^{l})\gamma^\rho\left(1-\gamma_5\right)\left[g_{\rho \alpha}-\frac{(p-k)_\rho (p-k)_\alpha}{M_W^2}\right]\nonumber\\
&\times& \left[g_{\nu \beta}-\frac{(p'-k)_\nu (p'-k)_\beta}{M_W^2}\right][g^{\beta\alpha}\left(2k-p^{}-p'^{}\right)^{\mu}\nonumber\\
&& +g_{}^{\alpha\mu}\left(2p^{}-p'-k\right)_{}^{\beta} +g_{}^{\mu\beta}\left(2p'^{}-k-p\right)_{}^{\alpha}] u(p),
\end{eqnarray}
since $\gamma_5$ anti-commutes with all $\gamma$ matrices on then the term with $\slashed{k}$ can be omitted and the $M_D^l$ remains,
\begin{eqnarray}
Numerator &=&
2M_{D}^{l}\bar{u}(p')\gamma^\nu\left(1+\gamma_5\right)\gamma^\rho\left[g_{\rho \alpha}-\frac{(p-k)_\rho (p-k)_\alpha}{M_W^2}\right]\nonumber\\
&\times& \left[g_{\nu \beta}-\frac{(p'-k)_\nu (p'-k)_\beta}{M_W^2}\right][g^{\beta\alpha}\left(2k-p^{}-p'^{}\right)^{\mu}\nonumber\\
&& +g_{}^{\alpha\mu}\left(2p^{}-p'-k\right)_{}^{\beta} +g_{}^{\mu\beta}\left(2p'^{}-k-p\right)_{}^{\alpha}] u(p).
\end{eqnarray}
After some computations and considering only the highest-order terms, it gives as
\begin{eqnarray}
&& Numerator =
\nonumber\\
&& 2M_{D}^{l}\bar{u}(p')\gamma^\nu\left(1+\gamma_5\right)\gamma_\alpha\left[ 
\delta^\alpha_\nu(2k-p-p')^\mu
+g^{\mu \alpha}(2p-p'-k)_\nu+\delta^\mu_\nu(2p'-k-p)^\alpha\right] u(p).
\end{eqnarray}
in terms of $l$ and using of the  identities (\ref{app2})
\begin{eqnarray}
Numerator &=&
2M_{D}^{l}\bar{u}(p')[\left(1-\gamma_5\right)(6zp^\mu+6yp'^\mu-3m_{\boldsymbol{\mu}}\gamma^\mu)+(1+\gamma_5)\gamma^\mu m_{\boldsymbol{\mu}}(y-2)] u(p).
\end{eqnarray}
We are only interested in the terms that contribute to the $\mu$AMM, and this can be achieved by employing the Gordon identity \ref{Gordon},
\begin{eqnarray}
	\delta\Gamma^{\mu }_{(\times)}&=&\frac{i}{8}M_{D}^{l}\,{\mathcal{G}_Rg_w}^{2}V^{{\boldsymbol{\mu}} l}_L\,V^{\dagger{\boldsymbol{\mu}} l}_R\int\frac{d^{4}l}{(2\pi)^4}\int_{0}^{1} \,dx\,dy\,dz\,\delta(x+y+z-1)\frac{6(zp^{\mu}+yp'^{\mu})}{(l^{2}-\Delta_{(a)})^{3}}
	.\label{gamm b}
\end{eqnarray}
Employing Eq. \eqref{app1}, one can easily find the following result
\begin{eqnarray}
	\delta\Gamma^{\mu }_{(\times)}&=&\frac{3g_w^{2}}{4(4\pi)^2}M_{D}^{l}\,\mathcal{G}_R V^{{\boldsymbol{\mu}} l}_L\,V^{\dagger{\boldsymbol{\mu}} l}_R\int_{0}^{1} \,dx\, \int_0^{1-x}\,dy\,\frac{(1-x-y)(p^{\mu}+p'^{\mu})}{\Delta_{(a)}}.
    \label{gamm b}
\end{eqnarray}
Working at the zero momentum transfer of the sterile neutrino yields
\begin{eqnarray}
	\delta\Gamma^{\mu }_{(\times)}&=&-\frac{3g_w^{2}}{8(4\pi)^2}M_{D}^{l}\,\mathcal{G}_R V^{{\boldsymbol{\mu}} l}_L\,V^{\dagger{\boldsymbol{\mu}} l}_R\int_{0}^{1} \,dx\,\,\frac{(1-x)^2(p^{\mu}+p'^{\mu})}{-x(1-x)m_{\boldsymbol{\mu}}^2+(1-x){M_W^2+xM_{D}^l}^2}\nonumber\\
  &=& -\frac{3g_w^{2}}{(16\pi)^2m_{\boldsymbol{\mu}}^4}M_{D}^{l}\,\mathcal{G} V^{{\boldsymbol{\mu}} l}_L\,V^{\dagger{\boldsymbol{\mu}} l}_R (p^{\mu}+p'^{\mu})\bigg\{
  -(m_{\boldsymbol{\mu}}^2-M_W^2+M_D^2)\ln{\frac{M_D^2}{M_W^2}}+2m_{\boldsymbol{\mu}}^2
  \nonumber\\&+&\frac{4m_{\boldsymbol{\mu}}^2M_D^2-2A}{\sqrt{A}}(\tan^{-1}(\frac{m_{\boldsymbol{\mu}}^2-M_W^2+M_D^2}{A})-\tan^{-1}(\frac{-m_{\boldsymbol{\mu}}^2-M_W^2+M_D^2}{A}))
  \bigg\} ,
    \label{gamm b33}
\end{eqnarray}
where  $A=-m_{\boldsymbol{\mu}}^4+2m_{\boldsymbol{\mu}}^2(M_W^2+M_D^2)-(M_W^2-M_D^2)^2$. Accordingly, the correction on the $\mu$AMM arising from the Sterile Neutrino at zero momentum transfer is given by
\begin{eqnarray}
a^{SN }_{(\times)}&=& -\frac{6g_w^{2}}{(16\pi)^2m_{\boldsymbol{\mu}}^3}M_{D}^{l}\,\mathcal{G}_R V^{{\boldsymbol{\mu}} l}_L\,V^{\dagger{\boldsymbol{\mu}} l}_R \bigg\{
-(m_{\boldsymbol{\mu}}^2-M_W^2+M_D^2)\ln{\frac{M_D^2}{M_W^2}}+2m_{\boldsymbol{\mu}}^2
\nonumber\\&+&\frac{4m_{\boldsymbol{\mu}}^2M_D^2-2A}{\sqrt{A}}(\tan^{-1}(\frac{m_{\boldsymbol{\mu}}^2-M_W^2+M_D^2}{A})-\tan^{-1}(\frac{-m_{\boldsymbol{\mu}}^2-M_W^2+M_D^2}{A}))
\bigg\}
.\label{agb}
\end{eqnarray}

\end{document}